\documentclass[pra,twocolumn,showpacs,superscriptaddress]{revtex4-1}
\usepackage{graphicx,amsmath}

\begin{document}

\title{Mechanism of entanglement preservation}
\author{Qing-Jun Tong}
\affiliation{Center for Interdisciplinary Studies, Lanzhou
University, Lanzhou 730000, China}
\affiliation{Department of Modern Physics, Lanzhou University, Lanzhou 730000, China}
\author{Jun-Hong An}
\email{anjhong@lzu.edu.cn}
\affiliation{Center for Interdisciplinary
Studies, Lanzhou University, Lanzhou 730000, China}
\affiliation{Department of Physics, National University of
Singapore, 3 Science Drive 2, Singapore 117543, Singapore}
\author{Hong-Gang Luo}
\affiliation{Center for Interdisciplinary Studies, Lanzhou University, Lanzhou 730000,
China}
\affiliation{ Key Laboratory for Magnetism and Magnetic Materials of MOE, Lanzhou University, Lanzhou 730000, China}
\author{C. H. Oh}\email{phyohch@nus.edu.sg}\affiliation{Department of Physics, National University of
Singapore, 3 Science Drive 2, Singapore 117543, Singapore}

\begin{abstract}
We study the entanglement preservation of two qubits locally
interacting with their reservoirs. We show that the existence of a
bound state of the qubit and its reservoir and the non-Markovian
effect are two essential ingredients and their interplay plays a
crucial role in preserving the entanglement in the steady state.
When the non-Markovian effect is neglected, the entanglement sudden
death (ESD) is reproduced. On the other hand, when the non-Markovian
is significantly strong but the bound state is absent, the
phenomenon of the ESD and its revival is recovered. Our formulation
presents a unified picture about the entanglement preservation and
provides a clear clue on how to preserve the entanglement in quantum
information processing.
\end{abstract}

\maketitle

\section{Introduction}
Entanglement is not only of fundamental interest to quantum
mechanics but also of great importance to quantum information
processing \cite{Nielsen00}. However, due to the inevitable
interaction of qubits with their environments, entanglement always
experiences degradation. Entanglement sudden death (ESD), a
phenomenon in which the entanglement between two qubits may
completely disappear in a finite time, has been predicted
theoretically \cite{Yu04} and subsequently been verified
experimentally \cite{Almeida07}, indicating specific behavior of
entanglement that differs from that of coherence. From the point of
view of applications, ESD is apparently disadvantageous to quantum
information processing.

Recently, Bellomo {\it et al.} \cite{Bellomo07} found that the
entanglement can revive after some time interval of ESD and thus
extends significantly the entangled time of the qubits. This
remarkable phenomenon, which has been experimentally observed
\cite{Xu09}, is physically due to the dynamical back action (that
is, the non-Markovian effect) of the memory environments
\cite{Bellomo07,Maniscalco08}. However, in many cases the finite
extension of the entangled time is not enough and thus it is desired
to preserve a significant fraction of the entanglement in the
longtime limit. Indeed, it was shown \cite{Bellomo08} that some
noticeable fraction of entanglement can be obtained by engineering
structured environment such as photonic band-gap materials
\cite{John1990, John1994}. According to these works, it is still
unclear whether the residual entanglement is fundamentally due to
the specific structured materials or due to certain physical
mechanisms. Is there any essential relationship between ESD and/or
its revival phenomena and the residual entanglement?

In this work we focus on these questions and elucidate the physical
nature of the residual entanglement. Before proceeding, it is
helpful to recall the physics of quantum electrodynamics of a single
two-level atom placed in a dielectric with a photonic band gap
\cite{John1990, John1994}. The coupling between the excited atom and
electromagnetic vacuum in the dielectric leads to a novel
photon-atom bound state \cite{John1994}, in which the fractional
atomic population on the excited state occurs, also known as
population trapping \cite{Lambropoulos00}. This result has been
verified experimentally for quantum dots embedded in a photonic
band-gap environment \cite{Lodahl04}. The population trapping has
been directly connected to the entanglement trapping due to the
structured environment \cite{Bellomo08}. Here we reveal that there
are two essential conditions needed to preserve the entanglement.
One is the existence of the bound state between the system and its
environment, which provides the ability to preserve the
entanglement, and the other is the non-Markovian effect, which
provides a way to preserve the entanglement. Our result can
reproduce ESD \cite{Yu04} when the non-Markovian effect is
neglected. The phenomenon of ESD and its revival discussed in Ref.
\cite{Bellomo07} results from the non-Markovian effect when the
bound state is not available. The interplay between the availability
of the bound state and the non-Markovian effect can lead to a
significant fraction of the entanglement preserved in the steady
state. We verify these results by considering two reservoirs modeled
by the super-Ohmic and Lorentzian spectra, respectively. The result
provides a general method on how to protect the entanglement by
engineering the environment.

This paper is organized as follows. In Sec. \ref{model}, the model
of two independent qubits in two reservoirs is introduced. By
exploring the eigen-spectrum of the model, we derive the condition
for the formation of bound states between the qubits and their
respective reservoirs and discuss its profound consequence on the
dynamics of the open two-qubit system. In Sec. \ref{etd}, the
entanglement preservation caused by the formation of a bound state
and the non-Markovian effect is studied numerically in different
cases of the environmental spectral density. Finally, we end with a
short conclusion and a discussion about the experimental feasibility
of our result in Sec. \ref{cd}.

\section{Model and decoherence dynamics}\label{model}
We consider a system consisting of two independent subsystems, each
of which contains a qubit coupled to a zero-temperature reservoir.
Due to the dynamical independence between the two subsystems, we can
first investigate the single subsystem, then extend our studies to
the double-system case. The Hamiltonian of each subsystem is
\cite{Scully97}
\begin{equation}
H=\omega _{0}\sigma _{+}\sigma _{-}+\sum_{k}\omega _{k}b_{k}^{\dag
}b_{k}+\sum_{k}(g_{k}\sigma _{+}b_{k}+g_{k}^{\ast }\sigma _{-}b_{k}^{\dag }),
\label{Hm}
\end{equation}
where $\sigma _{\pm }$ and $\omega _{0}$ are the inversion operators
and transition frequency of the qubit, $b_{k}^{\dag }$ and $b_{k}$
are the creation and annihilation operators of the $k$th mode with
frequency $\omega _{k}$ of the reservoir, and $g_k$ denotes the
coupling strength between the atom and the radiation field.

To check the spectrum of the Hamiltonian we first solve the
eigenvalue equation
\begin{equation}
H\left\vert \varphi _{E}\right\rangle = E\left\vert \varphi _{E}\right\rangle. \label{eig}
\end{equation}
If only one excitation is present in the system at zero temperature
initially, then $\left\vert \varphi _{E}\right\rangle
=c_{0}\left\vert +,\{0_{k}\}\right\rangle +\sum_{k}c_{k}\left\vert
-,1_{k}\right\rangle $, where $|\pm\rangle$ is the atomic excited
(or ground) state, and $|\{0_k\}\rangle$ and $|1_k\rangle$ are the
vacuum state and the state with only one photon in the $k$th mode of
the reservoir, respectively. Substituting Eq. (\ref{Hm}) into Eq.
(\ref{eig}), one has
\begin{equation}
y(E) \equiv \omega _{0}- \int_{0}^{\infty }\frac{J(\omega) }{\omega
- E}d\omega = E, \label{eigen}
\end{equation}
where $J(\omega)=\sum_k |g_k|^2\delta(\omega-\omega_k)$ is the
spectral density of the reservoir. The solution of Eq. (\ref{eigen})
highly depends on the explicit form of $J(\omega)$. If the reservoir
contains only one mode $\omega^\prime$, then $J(\omega) =
g^2\delta(\omega-\omega^{\prime })$. This is the ideal
Jaynes-Cummings (JC) model \cite{Scully97}, in which two bound
states in one excitation sector are formed and as a result the
dynamics of the system displays a lossless oscillation. When the
reservoir contains infinite modes, one can model $J(\omega)$ by some
typical spectrum functions such as the super-Ohmic or Lorentzian
forms.

We first consider the super-Ohmic spectrum $J(\omega)=\eta
\frac{\omega^3}{\omega_0^2}e^{-\omega/\omega_c}$, where $\eta$ is a
dimensionless coupling constant and $\omega_c$ characterizes the
frequency regime in which the power law is valid \cite{Weiss}. It
corresponds to the case in which the reservoir consists of a vacuum
radiation field, where $g_k\propto \sqrt{\omega_k}$ \cite{Scully97}.
The existence of a bound state requires that Eq. (\ref{eigen}) has
at least a real solution for $E < 0$. It is easy to check that the
solution always exists if the condition $y(0)<0$ (i.e. $ \omega
_{0}-2\eta \frac{\omega_c^3}{\omega_0^2}<0$) is satisfied.
Otherwise, no bound state exists. This condition can be fulfilled
easily by engineering the environment. For the Lorentzian spectrum
it is found that a criterion for the existence of a bound state
cannot be obtained analytically. In this case one can use the
diagrammatic technique shown later.

The existence of a bound state has a profound implication on the
dynamics of the single-qubit system such as the inhibition of
spontaneous emission \cite{John1990, John1994, Yablonovitch87}.
Furthermore, it also has an important impact on the entanglement
dynamics of a two-qubit system, which is governed by the master
equation \cite{Breuer02}
\begin{eqnarray}
\dot{\rho} (t) &=&\sum_{n=1}^2\{-i\Omega (t)[\sigma
_{+}^{n}\sigma_{-}^{n},\rho (t)]+\Gamma (t)[2\sigma _{-}^{n}\rho
(t)\sigma
_{+}^{n}  \notag \\
&&-\sigma _{+}^{n}\sigma _{-}^{n}\rho (t)-\rho (t)\sigma _{+}^{n}\sigma
_{-}^{n}]\},  \label{mstt}
\end{eqnarray}
where $\Omega (t)=-\text{Im}[\frac{\dot{c}_{0}(t)}{c_{0}(t)}]$, and
$\Gamma (t)=-\text{Re}[\frac{\dot{c}_{0}(t)}{c_{0}(t)}]$. It is shown that $c_0(t)$
satisfies
\begin{equation}
\dot{c}_{0}(t)+i\omega _{0}c_{0}(t)+\int_{0}^{t}c_{0}(\tau )f(t-\tau )d\tau
=0,  \label{c0}
\end{equation}
where $f(t-\tau )=\int_0^\infty J(\omega)e^{-i\omega(t-\tau
)}d\omega$. The time-dependent parameters $\Omega (t)$ and
$\Gamma(t)$ play the role of Lamb-shifted frequency and decay rate
of the qubits, respectively. The integro-differential equation
(\ref{c0}) contains the memory effect of the reservoir registered in
the time-nonlocal kernel function $f(t-\tau )$ and thus the dynamics
of the qubit displays a non-Markovian effect. If $f(t-\tau )$ is
replaced by the time-local function, then Eq. (\ref{mstt}) recovers
the conventional Born-Markovian master equation, where the
parameters become constants \cite{An092}, that is, $ \Gamma_0=\pi
J(\omega_0),~\Omega_0=\omega_0-P\int_0^\infty \frac{
J(\omega)d\omega}{\omega-\omega_0} $ with $P$ denoting the Cauchy
principal value.

The dynamical consequence of the bound state is decoherence
suppression \cite{Lambropoulos00}. If a bound state is absent, then
Eq. (\ref{eigen}) has only complex solutions. Physically this means
that the corresponding eigenstate experiences decay from the
imaginary part of the eigenvalue during the time evolution, which
causes the excited-state population to approach zero asymptotically
and the decoherence of the reduced qubit system. While if a bound
state is formed, then the population of the atomic excited state in
the bound state is constant in time because a bound state is
actually a stationary state with a vanishing decay rate during the
time evolution. So there will be some residual excited-state
population in the long-time limit. Due to this decoherence
suppression, we expect that the formation of a bound state plays a
constructive role in entanglement preservation under the
non-Markovian dynamics, as shown in the following section.

\section{Entanglement preservation }\label{etd}

To study the entanglement dynamics of the bipartite system, we use
the concurrence to quantify entanglement \cite{wootters98}. The
concurrence is
defined as $C(\rho )=\max \{0,\sqrt{\lambda _{1}}-\sqrt{\lambda _{2}}-\sqrt{%
\lambda _{3}}-\sqrt{\lambda _{4}}\}$, where the
decreasing-order-arranged quantities $\lambda _{i}$ are the
eigenvalues of the matrix $\rho (\sigma _{y}^{A}\otimes \sigma
_{y}^{B})\rho ^{\ast }(\sigma _{y}^{A}\otimes \sigma _{y}^{B})$.
Here $\rho ^{\ast }$ means the complex conjugation of $\rho $, and
$\sigma _{y}$ is the Pauli matrix. It can be proved that the
concurrence varies from 0 for a separable state to 1 for a maximally
entangled state. Consider an initially entangled state $ |\psi
(0)\rangle=\alpha \left\vert --\right\rangle +\beta \left\vert
++\right\rangle$, where $|\alpha |^{2}+|\beta |^{2}=1$. Then $C(t)$
can be calculated as $C(t) = \max\{0, Q(t)\}$, where
\begin{equation}
Q(t) = 2|\alpha\beta||c_0(t)|^2 - 2|\beta|^2 |c_0(t)|^2 [1 -
|c_0(t)|^2],  \label{t12}
\end{equation}
which indicates that the time-dependant factor of the excited state
population ($\left\vert c_{0}(t)\right\vert ^{2}$) determines solely
the entanglement dynamics.

\begin{figure}
\includegraphics[width = \columnwidth] {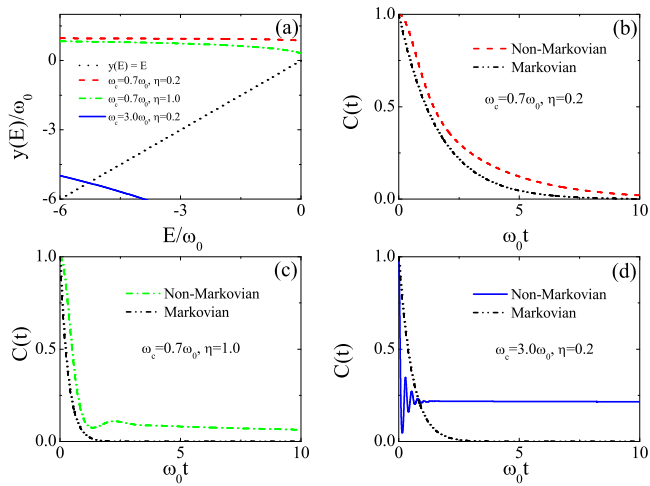}
\caption{(Color online) Entanglement dynamics of the two qubit
system with local super-Ohmic reservoirs. (a) Diagrammatic solutions
of Eq. (\ref{eigen}) with different parameters. $C(t)$ as a function
of time is shown in (b): $(\omega_c, \eta) = (0.7\omega_0, 0.2)$,
(c): $(\omega_c, \eta) = (0.7\omega_0, 1.0)$ and (d): $(\omega_c,
\eta) = (3.0\omega_0, 0.2)$. The parameter $\alpha$ is taken as
$0.7$. For comparison, $C(t)$ under the Markovian approximation has
also been presented by using the same parameters. } \label{fig1}
\end{figure}

\begin{figure}[tbp]
\includegraphics[width = 0.8\columnwidth]{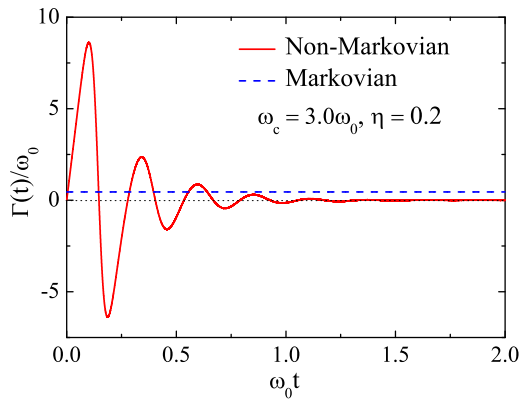}
\caption{(Color online) The decay rate $\Gamma(t)$ as a function of
time in the non-Markovian and Markovian cases. The parameters used
are $\omega_c = 3.0\omega_0$ and $\eta = 0.2$.}  \label{fig2}
\end{figure}

Now we are ready to study the entanglement dynamics of the two-qubit
system. Consider first the super-Ohmic case. Figure \ref{fig1} shows
the entanglement dynamics in different parameter regimes [i.e.,
$(\omega_c, \eta) = (0.7\omega_0, 0.2), (0.7\omega_0, 1.0)$ and
$(3.0\omega_0, 0.2)$]. For the first two parameter sets the bound
state is absent, while for the last one it is available, as shown in
Fig. \ref{fig1}(a). Whether the bound state exists or not plays a
key role in the entanglement preservation in the longtime limit.
When the bound state is absent, the residual entanglement approaches
zero in a long enough time, as shown by the non-Markovian lines in
Figs. \ref{fig1}(b) and \ref{fig1}(c). The difference between these
two cases is that Fig. \ref{fig1}(b) shows the weak-coupling regime,
where the non-Markovian effect is weak, while Fig. \ref{fig1}(c)
shows the strong-coupling regime, where the strong non-Markovian
effect leads to an obvious oscillation. When the bound state is
available, the situation is quite different, as shown in Fig.
\ref{fig1}(d). The entanglement first experiences some oscillations
due to the energy and/or information exchanging back and forth
between the qubit and its memory environment \cite{Pillo08}, then
approaches a definite value in the longtime limit, where the decay
rate approaches zero after some oscillations, as shown in Fig.
\ref{fig2}. {\it The entanglement preservation is a result of the
interplay between the existence of the bound state (providing the
ability to preserve the entanglement) and the non-Markovian effect
(providing a way to preserve the entanglement)}. The claim can be
further verified by the fact that the entanglement preservation is
absent in the Markovian dynamics, as shown by the Markovian lines in
Figs. \ref{fig1}(b), \ref{fig1}(c) and \ref{fig1}(d), where the
entanglement displays sudden death irrespective of the availability
of the bound state. This is because the Markovian environment has no
memory and the energy and information flowing from the qubit to its
environment is irreversible and the decay rate remains fixed (see
Fig. \ref{fig2}). In this case one has
\begin{equation}
C(t)=\max \{0,2e^{-2\Gamma_0t}\left\vert \beta \right\vert
[\left\vert \alpha \right\vert -\left\vert \beta \right\vert
(1-e^{-2\Gamma_0t})]\}, \label{cMar}
\end{equation}
which shows a finite disentanglement time when $|\alpha|<|\beta|$
\cite{Yu04}. In short, different to the need for a structured
environment as emphasized in Ref. \cite{Bellomo08}, our discussion
clearly reveals two essential conditions to preserve the
entanglement: the availability of the bound state and the
non-Markovian effect.

\begin{figure}[tbp]
\includegraphics[width = 0.8\columnwidth]{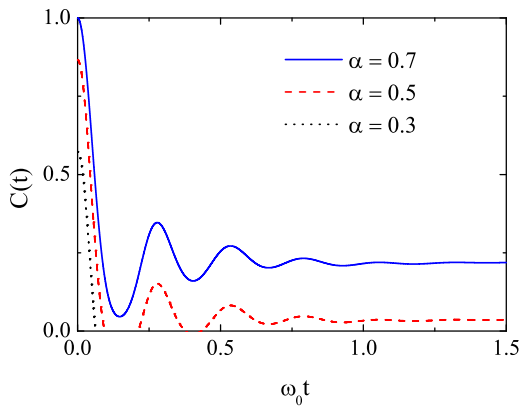}
\caption{(Color Online) The residual entanglement for different initial states with $\alpha = 0.7, 0.5$ and $0.3$. The other parameters used are the same as those in Fig. \ref{fig1}. }  \label{fig3}
\end{figure}

This discussion focused on an almost maximally entangled initial
state by taking $\alpha = 0.7$. In Fig. \ref{fig3} we show the
results for different initial states with different initial
entanglement. With decreasing initial entanglement, the residual
entanglement also decreases in the longtime limit and, finally, ESD
happens for $\alpha = 0.3$. The result can be understood from Eq.
(\ref{t12}). On the one hand, the residual entanglement is
determined by $c_0(\infty)$, which is directly related to the
property of the bound state. On the other hand, the residual
entanglement is also determined by the competition between the first
and the second terms in Eq. (\ref{t12}), which is dependent of the
initial state.

In order to make a comparative study and confirm our observations we
consider the Lorentzian spectrum if the reservoir is composed of a
lossy cavity,
\begin{equation}
J(\omega )=\frac{1}{2\pi }\frac{\gamma \lambda ^{2}}{(\omega
-\omega_0 )^{2}+\lambda ^{2}},  \label{t11}
\end{equation}%
where $\gamma $ is the coupling constant and $\lambda $ is the
spectrum width. This model has also been studied in Ref.
\cite{Bellomo07}, where the lower limit of the frequency integral in
$f(t-\tau)$ was extended from zero to negative infinity. This
extension is mathematically convenient but the availability of the
bound state is missed. Here we follow the original definition of the
frequency integral ranges.

\begin{figure}[h]
\includegraphics[width = \columnwidth] {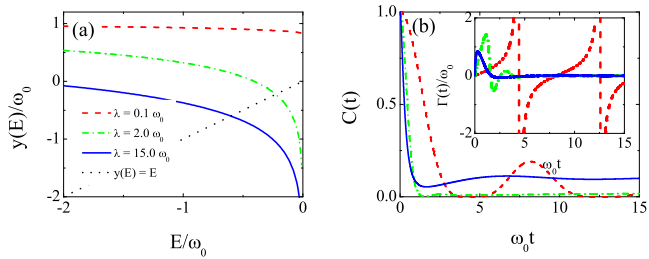}
\caption{(Color online) The entanglement dynamics with the
Lorentzian spectrum. (a) Diagrammatic solutions of Eq. (\ref{eigen})
with different parameters $\lambda = 0.1 \omega_0, 2.0\omega_0$ and
$15\omega_0$. (b) $C(t)$ as a function of time for the corresponding
three parameter regimes. The insert in (b) shows the decay rate as a
function of time. The other parameters used are $\gamma =
3.0\omega_0$ and $\alpha = 0.7$. } \label{fig4}
\end{figure}

\begin{figure}[h]
\includegraphics[width = \columnwidth] {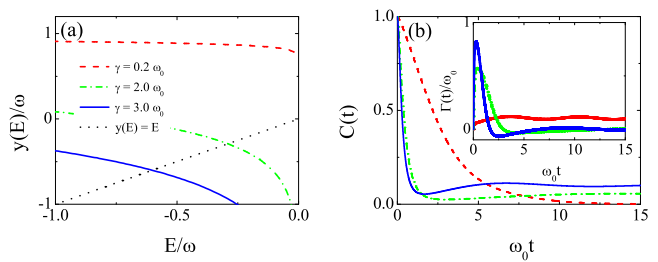}
\caption{(Color online) The same as Fig. \ref{fig4} but $\lambda =
15.0\omega_0$ is fixed and $\gamma = 0.2\omega_0, 2.0\omega_0$, and
$3.0\omega_0$.} \label{fig5}
\end{figure}

Our model with the Lorentzian spectral density corresponds exactly
to the extended damping J-C model \cite{Breuer02}. It is noted that
the strong coupling of the J-C model has been achieved in circuit
QED \cite{Wallraff04} and quantum dot \cite{Hennessy07} systems.
Figure \ref{fig4} shows the entanglement dynamics of the qubits
under the Lorentizan reservoir for different spectral widths in the
strong-coupling regime. When $\lambda = 0.1\omega_0$, Eq.
(\ref{eigen}) lacks the bound state. According to our discussion,
there is no residual entanglement in the longtime limit. This is
indeed true, as shown in Fig. \ref{fig4}(b). However, it is noted
that before becoming zero the entanglement exhibits ``sudden death"
and revives for several times. This is an analog of the central
result found in Ref. \cite{Bellomo07}, that is, the phenomenon of
ESD and revival. Apparently, this is due to the non-Markovian effect
with the revival being a result of back action of the memory
reservoir. The situation changes with increasing the spectral width,
and the bound states become available. A significant fraction of the
entanglement initially present is preserved in the longtime limit,
where the decay rates shown in the insert of Fig. \ref{fig4}(b)
approach zero in these cases. Likewise, the physical nature of the
entanglement preservation is still the interplay between the bound
state and the non-Markovian effect. The stronger the coupling is,
the more striking the entanglement oscillates as a function of time
and, consequently, the more noticeable the non-Markovian effect is,
as shown in Fig. \ref{fig5}. For $\gamma = 0.2 \omega_0$, the system
is in the weak-coupling regime, where the bound state is also not
available. As a result, ESD is reproduced in this case.

\begin{figure}[h]
\includegraphics[width = 0.8\columnwidth]{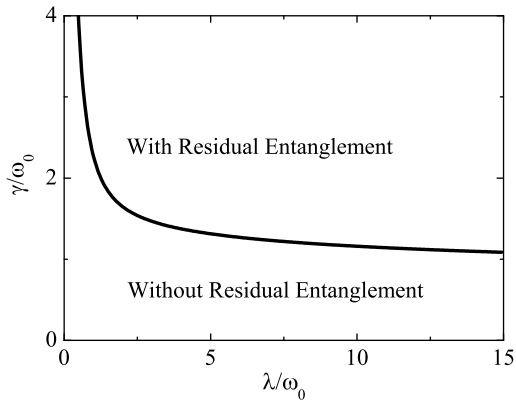}
\caption{The phase diagram of the residual entanglement in the
steady state for the Lorentzian spectrum.} \label{fig6}
\end{figure}

In Fig. \ref{fig6} we present a phase diagram of the entanglement in
the steady state for the Lorentzian spectral density. In the
large-$\gamma$ and small-$\lambda$ regime, the system approaches the
J-C model. In this situation the strong back action effect of the
reservoirs makes it difficult for the qubit system to form a steady
state. The entanglement oscillates with time but has no dissipation.
In the small-$\gamma$ and large-$\lambda$ regime, the non-Markovian
effect is extremely weak and our results reduce to the Markovian
case. In the limit of a flat spectral density, the Born-Markovian
approximation is applicable and the system has no bound state. This
is the case of ESD \cite{Yu04}.

\section{Conclusions and discussion}\label{cd}
In summary, we have studied the entanglement protection of two
qubits in two uncorrelated reservoirs. Two essential conditions for
preserving the entanglement are explored: the existence of the bound
state of the system and its reservoir and the non-Markovian effect.
The bound state provides the ability of the entanglement
preservation and the non-Markovian effect provides the way to
protect the entanglement. Previous results on the entanglement
dynamics in the literature can be considered as specific cases where
these two conditions have not been fulfilled at the same time. The
result here provides a unified picture for the entanglement dynamics
and gives a clear way on how to protect the entanglement. This is
quite significant in quantum information processing.

The presence of such entanglement preservation actually gives us an
active way to suppress decoherence. This could be achieved by
modifying the properties of the reservoir to form a bound state and
to approach the non-Markovian regime via the potential usage of
reservoir engineering \cite{Myatt00,Diehl08,Garraway}. Many
experimental platforms (e.g., mesoscopic ion traps \cite{Myatt00},
cold atom BECs \cite{Diehl08}, and the photonic crystal materials
\cite{Yablonovitch87} have exhibited the controllability of
decoherence behavior of relevant quantum systems through optimally
designing the size (i.e., modifying the spectrum) of the reservoir
and/or the coupling strength between the system and the reservoir.
It is also worth mentioning that a proposal aimed at simulating the
spin-Boson model, which is relevant to the one considered in this
paper, has been reported in a trapped ion system \cite{Porras}. On
the other hand many practical systems can now be engineered to show
the novel non-Markovian effect
\cite{Dubin07,Koppens07,Mogilevtsev08,Xu09,Galland08}. All these
achievements demonstrate that the recent advances have paved the way
to experimentally simulate the paradigmatic models of open quantum
system, which is one part of the newly emerging field of quantum
simulators \cite{Nori}. Our work sheds new light on the way to
indirectly control and manipulate the dynamics of a quantum system
in these experimental platforms. It provides a clue to preserving
the entanglement in quantum information processing.

\section*{ACKNOWLEDGMENTS}
This work is supported by the NSF of China under Grant No. 10604025,
the Gansu Provincial NSF of China under Grant No. 0803RJZA095, the
Program for NCET, and NUS A*STAR Research Grant No.
R-144-000-189-305.


\begin{thebibliography}{99}

\bibitem{Nielsen00} M. A. Nielsen and I. L. Chuang, \textit{Quantum
Computation and Quantum Information} (Cambridge University Press, Cambridge,
2000).

\bibitem{Yu04} T. Yu and J. H. Eberly, Phys. Rev. Lett. \textbf{93}, 140404
(2004); Science \textbf{323}, 598 (2009).

\bibitem{Almeida07} M. P. Almeida, F. de Melo, M. Hor-Meyll, A. Salles, S. P. Walborn, P. H. S. Ribeiro, and L. Davidovich,
Science \textbf{316}, 579 (2007).

\bibitem{Bellomo07} B. Bellomo, R. LoFranco, and G. Compagno, Phys. Rev.
Lett. \textbf{99}, 160502 (2007); Phys. Rev. A \textbf{77}, 032342
(2008).

\bibitem{Xu09} J.-S. Xu, C.-F. Li, M. Gong, X.-B. Zou, C.-H. Shi, G. Chen, and G.-C. Guo, Phys. Rev. Lett. {\bf 104}, 100502 (2010).

\bibitem{Maniscalco08} S. Maniscalco, F. Francica, R. L. Zaffino, N. Lo Gullo, and F. Plastina, Phys. Rev. Lett.
\textbf{100}, 090503 (2008).

\bibitem{Bellomo08} B. Bellomo, R. LoFranco, S. Maniscalco, and G.
Compagno, Phys. Rev. A 78 (2008) 060302(R).

\bibitem{John1990} S. John and J. Wang, Phys. Rev. Lett. {\bf 64}, 2418 (1990).

\bibitem{John1994} S. John and T. Quang, Phys. Rev. A {\bf 50}, 1764 (1994).

\bibitem{Lambropoulos00} P. Lambropoulos, G. M. Nikolopoulos, T. R. Nielsen, and S. Bay, Rep. Prog. Phys.
\textbf{63}, 455 (2000).

\bibitem{Lodahl04} P. Lodahl, A. F. van Driel, I. S. Nikolaev, A. Irman, K. Overgaag, D. Vanmaekelbergh, and W. L. Vos, Nature (London) \textbf{430} ,
654 (2004).

\bibitem{Scully97} M. O. Scully and M. S. Zubairy, \textit{Quantum Optics}
(Cambridge University Press, Cambridge, 1997).

\bibitem{Weiss} U. Weiss, {\it Quantum Dissipative Systems}, 3nd ed. (World Scientific, Singapore, 2008).

\bibitem{Yablonovitch87} E. Yablonovitch, Phys. Rev. Lett. \textbf{58}, 2059
(1987).

\bibitem{Breuer02} H.-P. Breuer and F. Petruccione, \textit{The Theory of
Open Quantum Systems} (Oxford University Press, Oxford, 2002).

\bibitem{An092} J.-H. An, Y. Yeo, and C. H. Oh, Ann. Phys. (N.Y.) {\bf 324},
1737 (2009).


\bibitem{wootters98} W. K. Wootters, Phys. Rev. Lett. \textbf{80}, 2245
(1998).

\bibitem{Pillo08} J. Piilo, S. Maniscalco, K. H\"{a}rk\"{o}nen, and K.-A.
Suominen, Phys. Rev. Lett. {\bf 100}, 180402 (2008).





%\bibitem{An09} J.-H. An \textit{et al.}, J. Phys. A: Math. Theor. \textbf{42}%
%, 015302 (2009).
\bibitem{Wallraff04} A. Wallraff, D. I. Schuster, A. Blais, L. Frunzio, R.- S. Huang,
J. Majer, S. Kumar, S. M. Girvin, and R. J. Schoelkopf, Nature (London)
{\bf 431}, 162 (2004).

\bibitem{Hennessy07} K. Hennessy, A. Badolato, M. Winger, D. Gerace, M. Atat\"{u}re, S. Gulde, S. F\"{a}lt, E. L. Hu, and A. Imamo\v{g}lu, Nature (London)
{\bf 445}, 896 (2007); A. Faraon, A. Majumdar, H. Kim, P. Petroff,
and J. Vu\v{c}kovi\'{c}, Phys. Rev. Lett. {\bf 104}, 047402 (2010).

\bibitem{Myatt00} C. J. Myatt, B. E. King, Q. A. Turchette, C. A. Sackett, D. Kielpinski, W. M. Itano, C. Monroe, and D. J. Wineland, Nature
(London) {\bf 403}, 269 (2000); Q. A. Turchette, C. J. Myatt, B. E.
King, C. A. Sackett, D. Kielpinski, W. M. Itano, C. Monroe, and D.
J. Wineland, Phys. Rev. A {\bf 62}, 053807 (2000).

\bibitem{Diehl08}S. Diehl, A. Micheli, A. Kantian, B. Kraus, H. P. B\"{u}chler,
and P. Zoller, Nat. Phys. {\bf 4}, 878 (2008).

\bibitem{Garraway} I. E. Linington and B. M. Garraway, Phys. Rev. A
{\bf 77}, 033831 (2008).


\bibitem{Porras} D. Porras, F. Marquardt, J. von Delft, and J. I.
Cirac, Phys. Rev. A {\bf 78}, 010101(R) (2008).

\bibitem{Dubin07} F. Dubin, D. Rotter, M. Mukherjee, C. Russo, J. Eschner, and R. Blatt, Phys. Rev. Lett. \textbf{98}, 183003 (2007).

\bibitem{Koppens07} F. H. L. Koppens, D. Klauser, W. A. Coish, K. C. Nowack, L. P. Kouwenhoven, D. Loss, and L. M. K. Vandersypen, Phys. Rev. Lett.
\textbf{99}, 106803 (2007).

\bibitem{Mogilevtsev08} D. Mogilevtsev, A. P. Nisovtsev, S. Kilin, S. B. Cavalcanti, H. S. Brandi, and L. E. Oliveira, Phys. Rev. Lett. \textbf{100},
017401 (2008).

\bibitem{Galland08} C. Galland, A. H\"{o}gele, H. E. T\"{u}reci, and A.
Imamo\v{g}lu, Phys. Rev. Lett. {\bf 101}, 067402 (2008).


\bibitem{Nori} I. Buluta and F. Nori, Science {\bf 326}, 108 (2009).
%\bibitem{Miyamoto05} M. Miyamoto, Phys. Rev. A \textbf{72,} 063405 (2005).







%\bibitem{Hu92} B. L. Hu, J. P. Paz, and Y. Zhang, Phys. Rev. D \textbf{45},
%2843 (1992).




\end{thebibliography}
\end{document}